# On the peer review reports: It's not the size that matters ... really?

Abdelghani Maddi[1] and Luis Miotti[2]

[1] *abdelghani.maddi@cnrs.fr*
GEMASS – CNRS – Sorbonne University, 59/61 rue Pouchet 75017 Paris, France

[2] *egidio.miotti@hceres.fr*
Science and Technology Observatory (Hcéres), 2 rue Albert Einstein 75013 Paris, France

**Abstract**
Scientometers and sociologists of science have spilled much ink on the topic of peer review over the past twenty years, given its primordial role in a context marked by the exponential growth of scientific production and the proliferation of predatory journals. Although the topic is addressed under different prisms, few studies have empirically analyzed to what extent it can affect the quality of publications. Here we study the link between the length of reviewers' reports and the citations received by publications. To do this, we used data from the Publons database (58,093 peer review reports). We have adjusted this sample to match the WoS database structure. Our regression results show that peer review positively affects the quality of publications. In other words, the more in-depth (longer) the referees' reports are, the greater the publication improvements will be, resulting in an increase in citations received. This result is important from both the point of view of reviewers and that of journal's chiefs-editors. Even if it is not a remunerated activity, it is important that it be more valued at least within the framework of research evaluation exercises, given its positive impact on science.

**Introduction**
The number of scientific journals has grown exponentially over the past twenty years that it is "*now impossible for a researcher to apprehend all of the published literature in his field*" (Larivière, 2017). This provides fertile ground for the development of predators seeking to take advantage of the situation. The number of predatory journals is currently estimated at 13,000 journals with more than hundred thousand published articles (Boukacem-Zeghmouri et al., 2020; C. Shen & Björk, 2015).

(Boukacem-Zeghmouri et al., 2020) provided a detailed description of the predation phenomenon in the scholarly publishing. Predatory journals are characterized by three main strategies: all predatory journals are Open Access (OA) based on a "author-pays" business model, in which peer review is not rigorous (or even absent) and displays "standard markers" of scientific publications (e.g. ISSN, DOI, etc.).

As much of the literature has shown, the publishing world cannot be categorized solely as black or white (Cobey et al., 2019; Siler, 2020). Sometimes the line between what a predatory journal and a real scientific journal is not clear. One of the criteria that allows deciding is the quality of the peer review practiced within journals. In particular if, all the other administrative/editorial characteristics surrounding the publication are not false or usurped (e.g. ISSN, DOI). This is the case, for example, of newly created journals, little known, with low bibliometric indicators (e.g. low impact factor). To grow and exist in the current oligopolistic market of scholarly communication, with many barriers to entry, these journals sometimes adopt aggressive commercial strategies. Therefore, they may look like predatory journals without actually being. This issue creates a certain anxiety among the scientific community who fear for their reputation if they happen to publish in journals, which prove to be predatory, or without scientific basis (Frandsen, 2019; Kolata, 2017). Some have described a kind of *omerta* within the scientific community (Djuric, 2015). In this regard, (Björk & Solomon, 2012) stressed that the OA status of journals should not be a reason for distrust for researchers, since not all OA journals are predatory. (Björk and Solomon, 2012) have also emphasized that researchers should, on the other hand, carefully check the quality standards of journals before submitting their work.

In this context, the role of peer review is paramount and deserves special attention. How can we "assess" peer review? Does the length of reviewers' reports improve the quality of manuscripts? Alternatively, is it only a rhetorical instrument intended for publishers and authors? Do reviewers write detailed reports just to please editors? The analysis carried out by Publons team is the only one, to our knowledge, that has investigated the link between the size of reviewers' reports and the citation impact on a large sample of over 378,000. (Publons, 2018) showed the existence of a positive correlation between the average number of words in evaluation reports and the impact of journals. Nevertheless, the study concluded that it is not possible to say with certainty that longer reviews are better or worse than shorter ones: "*Great reviews can be short and concise. Poor reviews can be long and in-depth, or vice-versa*". The analysis carried out by Publons remains quite descriptive and does not indeed allow confirming the positive link observed between the two variables. For this, an econometric model is necessary to neutralize the effects of the variables affecting citation impact (especially the journal's prominence and prestige, or open access availability) on the one hand, and the length of the reviewers' reports on the other.

In this study, we hypothesize that reviewers' reports actually improve the quality of publications. The interest they have received within the scientific community and results in a high number of citations reflects to some extent their quality (proxy). Thus, the length of reviewers' reports is representative of the number of modifications/improvements that authors must make to their manuscript for an eventual acceptance for publication. In other words, there should be a positive relationship between the length of reviewers' reports and the quality of scientific publications.

To test this hypothesis, we use on the one hand the data provided by Publons (https://publons.com/) for the length of reviewer' comments, and on the other hand on the data from the Web of Science (WoS) database for the calculation of bibliometric indicators. We performed several econometric models to analyze the link between the two variables: size of reviewers' reports and citation impact.

The article's plan is the following: First, we present a literature review on the link between peer review (in general) and the quality of publications. Then, we present some descriptive statistics on the database used (Publons), followed by the results section. Finally, the conclusion and the discussion return to the results obtained and their scope and implications.

**Literature review**
Peer review is a topic that has garnered much ink in the literature in Scientometrics and the sociology of science. We will only present here the work related to our research question, which is the impact of the peer review process on the quality of publications, approached by citations indicators.

Peer review processes are generally long, with disparities across disciplines (Björk & Solomon, 2013; Cornelius, 2012; Huisman & Smits, 2017; Kareiva et al., 2002). Especially for accepted publications (Bilalli et al., 2020). Over the past ten years, the duration has lengthened because of the inflation in the number of publications submitted to journals. On the other hand, the acceptance rate has increased by around 50% (Björk, 2018). In addition to the constraints related to the number of publications submitted each week to journals, the fact that referees are not paid for their review activity can contribute to slowing down the process (Azar, 2007;

Moizer, 2009; Toroser D et al., 2016). The increase in the technicality of publications can also increase the evaluation time in certain disciplines (Azar, 2007).

Whatever the reasons that may affect the evaluation times of scientific publications, the final objectives of peer review are multiple and revolve around improving the effective quality of publications. Chataway & Severin (2020) outlined eight purposes of peer review: (1) assess the contribution and originality of a manuscript, (2) perform quality control, (3) improve manuscripts, (4) assess the suitability of manuscripts for the topics of the journal, (5) provide a decision-making tool for editors, (6) provide comments and peer feedback, (7) strengthen the organization from the scientific community, and (8) provide some sort of accreditation for published papers.

Studies that assess the quality of the review process, from the perspective of authors or editors, are quite rare. Huisman & Smits (2017) analyzed data from 3,500 experiences and author reviews published on SciRev.sc (https://scirev.org/). The SciRev.sc interface groups together author reviews and ratings assigned to journals based on the quality (perceived by them) of the review process. Huisman & Smits (2017) have shown that, unsurprisingly, reviews with short response times tend to score higher. The same goes for experiences that resulted in a positive response from the journal. Drvenica et al. (2019) arrived at similar results on a sample of 193 authors. Furthermore, Huisman & Smits (2017) have shown that journals in disciplines where the evaluation processes are relatively long are on average better rated than journals in disciplines where the processes are short. In a more recent study, Pranić et al. (2020) analyzed the perception of authors and editors of the quality of peer reviews in 12 journals: 809 manuscripts and 313 opinions and recommendations. Pranić et al. (2020) found that authors give high ratings and positive perception when reviewers' comments recommend acceptance, unlike comments that recommend rejection (which is to be expected). On the other hand, the evaluations recommending a revision, are of better quality according to the indicator used (*Review Quality Instrument* - RQI). In addition, Pranić et al. (2020) have shown a strong association between the recommendation of referees and the publication decision of editors.

Studies that cross the length of the peer review process with the impact of journals or publications are even rarer. The study of Pautasso & Schäfer (2010) on 22 ecological / interdisciplinary journals, showed the existence of an inverse relationship between the acceptance delay and the impact factor of the journals. These same journals, however, have relatively low acceptance rates. The analysis of Shen et al. (2015) on the three journals (Nature, Science and Cell), over the period 2005-09, arrived at different conclusions from those of Pautasso & Schäfer (2010). For the three journals studied, the authors observed an inverse relationship between editorial time and the number of citations received.

There is room for improvement in the existing literature on assessing the length of the peer review process. The analyzes of Huisman & Smits (2017), Drvenica et al. (2019) and Pranić et al. (2020) can be criticized insofar as they assess the perceived quality of the evaluation and not the intrinsic quality. The perceived quality depends very much on the final decision and the length of the revisions requested by the referees; it is therefore very subjective. Hence, a thorough analysis of referees can result in a lot of criticism and be a source of improvement of the publications, even if the authors perceive it negatively, because it delays the publication (and adds a lot of additional work). To measure intrinsic quality, it is essential to cross-reference the performance indicators (eg citations received) of journals or articles.

Likewise, the study by Pautasso & Schäfer (2010) which showed that high impact journals have short turnaround times ignore the type of decision made by the journals. In general, the number of articles submitted to high-impact journals is important, as is the rejection rate. The rejection decision usually comes a few days / weeks after submission, reducing the average processing time for the entire review. Taking into account the time to get the first response and the type of response are important elements in the analysis. Thus, it can be assumed that lead times for accepted papers are higher in high-impact journals.

**Data**

*Web of Science data*
The data about citations scores and disciplinary assignation of publications has been extracted from the "Observatoire des Sciences et Techniques" (OST) in-house database. It includes five indexes of WoS available from Clarivate Analytics (SCIE, SSCI, AHCI, CPCI-SSH and CPCI-S. for more information see: https://clarivate.com/webofsciencegroup/solutions/webofscience-platform/) and corresponds to WoS content indexed through the end of November 2020. We have limited the analysis only to the original contributions; i.e. the following documents types: "Article", "Conference proceedings" and "Review".

*Publons data*
Publons is an interface created in 2012 specializing in peer review issues. In 2017, the supplier of the WoS database, *Clarivate Analytics*, acquired Publons. Currently, this database index over 2 million researcher profiles who share their peer review experiences. Publons offers a free service for researchers to promote their contributions. Publons also made available more than 300,000 items designating the characteristics of referee evaluation reports in journals: average word count, country of referee and journal. As part of this study, Publons made its database available to us by adding article identifiers (WoS UT). This allowed us to easily match Publons data with that of the WoS database and add other variables. Our final dataset contains 58,093 distinct publications reviewed by 62,152 reviewers (table 1). Publons database provides word count per review.

**Table 1: Number of reports per publication**

| #Reports by publication | #Publications | Freq. (%) | #Reports | Freq. (%) |
|---|---|---|---|---|
| 1 | 54,296 | 93.464 | 54,296 | 87.360 |
| 2 | 3,552 | 6.114 | 7,104 | 11.430 |
| 3 | 229 | 0.394 | 687 | 1.105 |
| 4 | 15 | 0.026 | 60 | 0.097 |
| 5 | 1 | 0.002 | 5 | 0.008 |
| **Total** | 58,093 | 100.0 | 62,152 | 100.0 |

**Method**

*Adjustment of Publons data*

The data extracted from Publons database represents a sample of the global WoS database, which we consider in this work as the population. When working with samples (e.g. public opinion polls), it is important to adjust data for population parameters. Thus, to study whether the conclusions drawn on the Publons database can be extended to the WoS database, we analyzed whether the respective structures are similar, based on a certain number of variables considered significant in explaining citation scores. Given that there are statistically significant differences between the Publons structure and that of the WoS, we adjusted the Publons sample using the procedure called "raking ratio"(Deming & Stephan, 1940; Deville et al., 1993), which provides adjustment weights. The variables used for the adjustment of the Publons sample:

- Publication year: dummy variables;
- Open access: binary variable (Yes/No);
- Financing: one, two, three, four or more;
- Number of countries: one, two, three, four or more;
- Scientific classification: ERC codes (dummy variable).
- The impact of journal (5 classes : <0.8, [0.8 , 1.2[, [1.2 , 1.8[, [1.8 , 2.2[, >=2.2), for the calculation method see: (Maddi & Sapinho, 2022).

Figure 1 summarizes the Publons data adjustment procedure. Appendix A.1 and A.2 show the structure of the Publons sample and of the relevant WoS population, before and after adjustment.

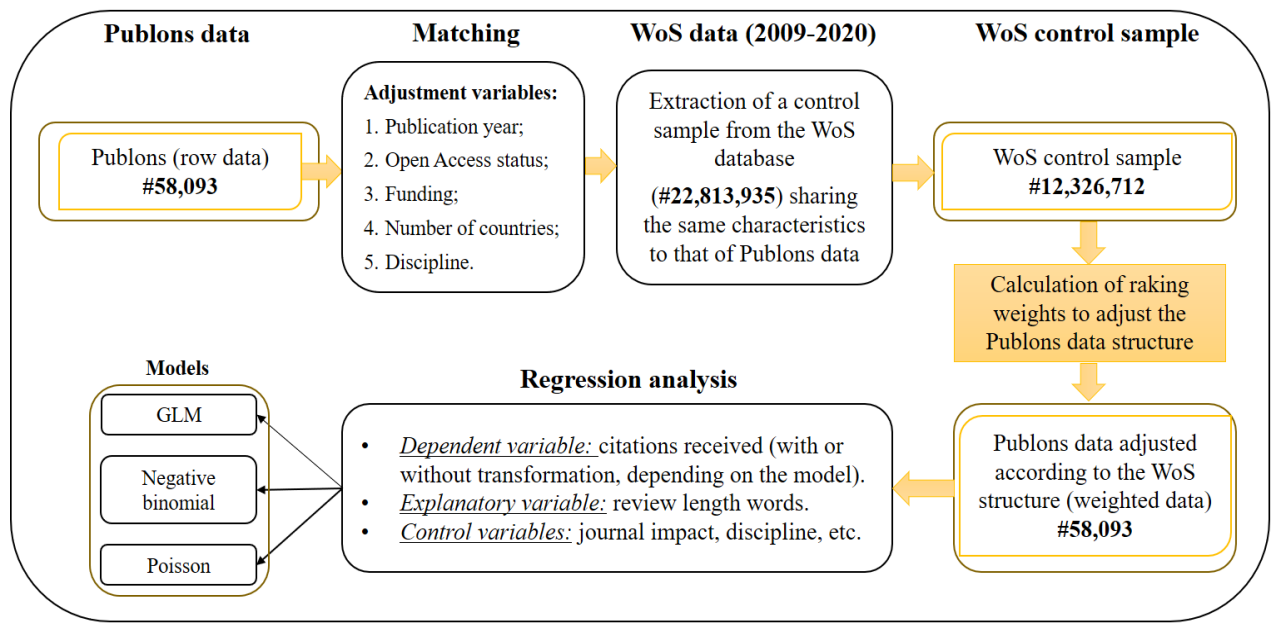

Figure 1: Publons data adjustment method

*Regression analysis*

Before performing the regressions, some processing is necessary to harmonize the data. To simplify the statistical and econometric analysis, given the fact that nearly 93.5% of publications present a single report (see Table 1), we proceeded to a random draw to build a dataset with a single reviewer' report per document. This solution seems optimal because the Publons database does not provide the number of reviewers (number of reports) by document and, consequently, the use of procedures such as the clustering of residues seems to be not necessary in this case.

**Table 2: descriptive statistics for intervals (review length words)**

| Lower bound | Upper bound | Effectif | Freq. |
|---|---|---|---|
| *0* | *200* | *20,939* | *36,0%* |
| 200 | 400 | 14,141 | 24.3% |
| 400 | 600 | 9,160 | 15.8% |
| 600 | 800 | 5,334 | 9.2% |
| 800 | 1,000 | 3,162 | 5.4% |
| 1,000 | 1,200 | 1,998 | 3.4% |
| 1,200 | 1,400 | 1,137 | 2.0% |
| 1,400 | 1,600 | 701 | 1.2% |
| 1,600 | 1,800 | 427 | 0.7% |
| 1,800 | 2,000 | 317 | 0.5% |
| 2,000 | 2,200 | 218 | 0.4% |
| 2,200 | 2,400 | 144 | 0.2% |
| 2,400 | 2,600 | 106 | 0.2% |
| 2,600 | 2,800 | 69 | 0.1% |
| 2,800 | 3,000 | 58 | 0.1% |
| *3,000* | *13,671* | *182* | *0.3%* |
| **Total** | | **58,093** | **100.0%** |

| | |
|---|---|
| *Number of observations - total* | 58,093 |
| *Number of observations - sample* | 36,972 |
| *Average - total* | 431 |
| *Average - sample* | 610 |
| *Median - total* | 306 |
| *Median - sample* | 482 |

With regard to the distribution of the number of words in the reports, the analysis by percentiles shows that just over 60% of the reports are less than 400 words (see table 2). It should be noted that approximately 10% of the total, or 5174 reports, have a report length of less than or equal to 20 words (which does not correspond to a normal length of a reference report). At the same time, 182 reports exceed 3,000 words (up to a maximum of 13,671 words), which represents approximately 0.3% of the total. We excluded from the regressions these two extremes of the distribution adding noise to the data. A preliminary econometric analysis shows the total lack of significance of the coefficient associated with the length of reports on any citation measure used, when the number of words is less than 200. Consequently, in the following, the econometric analysis will relate to the slices between 200 and 3000 words, which represents a little more than 64% of the reports. Table 3 shows the variables used for the regressions.

**Table 3: Model variables by type**

| Type | Variable | Designation |
|---|---|---|
| **Dependent variable** | Citation Score | Depending on the model, we used the citation score with or without transformation as follows:<br>- the number of citations (without window), expressed in raw numbers,<br>- using a logarithmic transformation,<br>- using a square root,<br>- Finally the logarithm of the Normalized Citation Score (NCS: see (Waltman et al., 2011) for calculation method). |
| **Explanatory variable** | Review length words | This is the word count per review report per publication. We tested two intervals: 200 to 3000 words, and 500 to 3000 words. |
| **Control variables** | Journal impact factor | The classic two-year impact factor. |
| | Open Access | The status of free access Yes / No (with "No" as a reference value). |
| | Funding | The number of funding received by the publication (We used for this the WoS acknowledgment field). It is expressed in numbers with five binary modalities:<br>- Without funding (Yes / No),<br>- *One funding (Yes / No), (reference value)*<br>- Two funding (Yes / No),<br>- Three funding (Yes / No),<br>- Four or more funding (Yes / No). |
| | Scientific collaboration | The number of countries involved in the publication (using author addresses). It is expressed in numbers with four binary modalities:<br>- A country (Yes / No),<br>- *Two countries (Yes / No), (reference value)*<br>- Three countries (Yes / No),<br>- Four or more countries (Yes / No). |
| | Discipline | 27+2 disciplinary levels, according to the nomenclature used by the OST based on the WoS subjects categories and ERC panels. For more detail, see: https://figshare.com/articles/dataset/OST_-_Classification_of_WoS_subject_categories_into_27_2_ERC_panels_/21707543 |
| | Publication year | The publication year, between 2010 and 2020 |

**Results**

Table 4 shows the results of the estimates by model and by range of the word count of the evaluation reports. As can be seen on the value and significance of the coefficients, we find the same trends as those shown in the literature, concerning the control variables. Thus, open access has a positive impact on the citations received, as well as funding and international cooperation.

What interests us in this paper is the value and significance of our explanatory variable; the number of words per reviewer's report. Of the nine estimation models, eight indicate, ceteris paribus, the existence of a positive and statistically significant impact of the word count of reviewers' reports on the citations received by publications. This result tends to confirm our hypothesis that peer review contributes to improving the quality of publications. In other words, the number of words in the reports, which we used as a proxy for the modifications requested by the referees, improves the quality of the publications. The citations that the publications receive reflect this quality improvement, as here we consider the citations as a proxy for the interest aroused by the publication within the scientific community.

However, we should take this result with some caution for several reasons. If the correction of the structural bias allows a comparison of the Publons database with the control sample taken from WoS, it does not eliminate the selection bias. This selection bias appears if the publication of the reviewers' reports in Publons is not done randomly but meets specific criteria (personal -age, gender-, and professionals -seniority, affiliation, scientific profile-, journal constraints concerning obligation/choice to publish reports, etc.). A few leads suggest the existence of a selection bias:

- The Open Access rate in Publons is 43.3% (and 47.2% in the case of a sample limited to reports with a word count in the 200-3000 range) and in the control sample drawn of WoS is 28.7%. This over-representation in Publons may indicate non-random behavior in reviewer reporting. Therefore, it is not impossible that reviewers, who dump their entire reports in Publons, can do so strategically. In other words, we only put on Publons well-produced reports. A recent paper (Radzvilas et al., 2022), based on evolutionary game theory, indicates that the type of peer review of the journal directly affects the behavior of researchers and reviewers. Specifically, the paper indicates that if the journal practices open peer review, it encourages researchers and reviewers to put in more effort (because it directly affects their reputation). Our results seem to indicate this phenomenon (especially on the side of the reviewers) and merit further investigation. Another paper (Bravo et al., 2019) shows that the publication of reports can change the tone used by reviewers (especially younger ones) due in some cases to fears of reprisals.
- The funding rate in Publons is 73.7% (and 76.3% in the case of a sample limited to reports with word counts in the 200-3000 range) and in the WOS control sample is of 53.9%. We know that the impact of funding on the citations of articles is positive and significant and, therefore, to hypothesize better quality. This could prompt some reviewers to declare their reports to Publons.

In sum, our regression results are both interesting and intriguing with respect to the Publons data, as they leave many questions unanswered. To be able to generalize our conclusions, it would be interesting to do the same work on a random sample of peer review reports. One idea would be to carry out the analysis using a panel of journals that practice open peer review, while respecting the problem of representativeness of the data and their structure in relation to the population (WoS, Scopus or any other large database - e.g. OpenAlex).

At this stage, we can simply say that the preliminary results suggest the existence of a positive link between the length of the peer evaluation reports and the quality of the publications (approximated by the citations).

| Model | GLM (identity) | GLM (identity) | GLM (identity) | GLM (Poisson) | GLM (Poisson) | GLM (Negative binomial) | GLM (Negative binomial) | GLM (identity) | GLM (identity) |
|---|---|---|---|---|---|---|---|---|---|
| Words | 200 - 3000 | 200 - 3000 | 500 - 3000 | 200 - 3000 | 500 - 3000 | 200 - 3000 | 500 - 3000 | 200 - 3000 | 500 - 3000 |
| | (1) | (2) | (3) | (4) | (5) | (6) | (7) | (8) | (9) |
| VARIABLES | Citations (Square root) | Citations (log) | Citations (log) | Citations | Citations | Citations | Citations | Score NCS (log) | Score NCS (log) |
| Open Access | 0.132*** | 0.0489*** | 0.0600** | 0.115*** | 0.125** | 0.0545** | 0.0634* | 0.0101 | 0.0231* |
| | (0.0422) | (0.0165) | (0.0249) | (0.0405) | (0.0576) | (0.0224) | (0.0325) | (0.00821) | (0.0126) |
| Without financing | 0.110** | 0.00561 | -0.0118 | 0.0893 | 0.0633 | 0.0606** | 0.0368 | 0.0110 | 0.00426 |
| | (0.0531) | (0.0221) | (0.0343) | (0.0544) | (0.0825) | (0.0306) | (0.0443) | (0.0110) | (0.0168) |
| Two financing | 0.0810 | 0.0534** | 0.0569* | 0.00891 | -0.0354 | 0.0352 | 0.0282 | 0.0192* | 0.0203 |
| | (0.0550) | (0.0228) | (0.0344) | (0.0465) | (0.0649) | (0.0294) | (0.0441) | (0.0110) | (0.0167) |
| Three financing | 0.113* | 0.0873*** | 0.0636 | -0.0321 | -0.0909 | 0.0348 | -0.00661 | 0.0397*** | 0.0258 |
| | (0.0604) | (0.0256) | (0.0388) | (0.0463) | (0.0655) | (0.0330) | (0.0445) | (0.0128) | (0.0182) |
| Four or more financing | 0.283*** | 0.141*** | 0.140*** | 0.0567 | 0.0379 | 0.108*** | 0.105** | 0.0685*** | 0.0658*** |
| | (0.0594) | (0.0229) | (0.0334) | (0.0497) | (0.0728) | (0.0313) | (0.0450) | (0.0117) | (0.0169) |
| One country | -0.236*** | -0.0951*** | -0.126*** | -0.222** | -0.272** | -0.153*** | -0.208*** | -0.0534*** | -0.0640*** |
| | (0.0694) | (0.0198) | (0.0294) | (0.0905) | (0.128) | (0.0434) | (0.0577) | (0.0114) | (0.0165) |
| Three countries | 0.287 | 0.115*** | 0.0824 | 0.0184 | -0.0780 | 0.0445 | -0.0519 | 0.0710** | 0.0643 |
| | (0.197) | (0.0413) | (0.0588) | (0.157) | (0.228) | (0.0677) | (0.0870) | (0.0294) | (0.0444) |
| Four or more countries | 0.653** | 0.258*** | 0.351*** | 0.221 | 0.343 | 0.352* | 0.473 | 0.170*** | 0.207** |
| | (0.273) | (0.0801) | (0.126) | (0.181) | (0.223) | (0.193) | (0.288) | (0.0536) | (0.0845) |
| **REVIEW LENGTH WORDS (log)** | **0.0754**** | **0.0173** | **0.0898**** | **0.107**** | **0.253**** | **0.0392*** | **0.103**** | **0.0144*** | **0.0501**** |
| | **(0.0381)** | **(0.0145)** | **(0.0305)** | **(0.0434)** | **(0.0646)** | **(0.0209)** | **(0.0403)** | **(0.00773)** | **(0.0171)** |
| Journal impact factor (log) | 1.507*** | 0.718*** | 0.732*** | 0.964*** | 0.978*** | 0.902*** | 0.912*** | 0.244*** | 0.250*** |
| | (0.0434) | (0.0149) | (0.0221) | (0.0333) | (0.0440) | (0.0208) | (0.0303) | (0.00835) | (0.0130) |
| Constant | 4.681*** | 2.797*** | 2.401*** | 3.044*** | 2.304*** | 3.256*** | 2.996*** | 0.591*** | 0.380*** |
| | (0.302) | (0.113) | (0.225) | (0.219) | (0.446) | (0.159) | (0.329) | (0.0569) | (0.121) |
| Observations | 36,972 | 36,972 | 17,659 | 36,972 | 17,659 | 36,972 | 17,659 | 36,972 | 17,659 |

Robust standard errors in parentheses
*** p<0.01, ** p<0.05, * p<0.1

*Note: Controlled by year of publication and scientific class (28)*

**Table 4: Results of estimates of the impact of the length of reviewers' reports on citations**

## Conclusion and discussion

Through this paper, we have sought to analyze empirically to what extent peer review improves the quality of publications. To do this, we used data from Publons (#85,093) with a structure adjusted to that of the WoS database (control group of 12,326,712 publications). The proxy used for peer review is the length of review reports expressed in number of words (provided by Publons). On the publications side, the proxy used to measure the quality (interest aroused within the scientific community) of the publications are the citations received. We carried out several regression models, each time putting the citation scores as the explained variable, and the length of the reviewers' reports as the explanatory variable (while adding a certain number of control variables).

In addition to the "classic" results on the importance of Open Access, funding or international collaboration on citations, we have shown the existence of a strong and statistically significant relationship between the length of reviewers' reports and citations received. The longer the referees' reports, the more the citations tend to increase. This result is consistent with our hypothesis formulated in the introduction, which consists in saying that peer review fulfills its function of improving scientific publications. In other words, long referees' reports go hand in hand with the number of modifications / improvements requested, and is a source of quality improvement, which is seen through the citations received.

This result is important from both the point of view of reviewers and that of journal's chiefs-editors. Thus, by affecting the intrinsic quality of publications, peer review work is an integral part of the creation of new knowledge and the development of science. Even if it is not a remunerated activity, it is important that it be more valued at least within the framework of research evaluation exercises, given its positive impact on science.

From the editors' point of view, this result underlines that a good peer review report is generally long (even if we can be short and concise), and therefore can take time. On this point, it would be interesting to analyze (1) from what length of the reports that the impact on citations becomes significant, and (2) if the impact increases (regression coefficient) as the average number of words increases. This analysis will be the subject of future improvements in this contribution.

## Limitations

Despite the effort to correct the data structure of the Publons database, there are still some limitations that cannot be addressed at this time. This is a base built from the volunteering of researchers to disseminate (promote) their reviewing activities. As a result, the data is not necessarily representative of all the evaluation reports due to this selection bias. Therefore, it would be interesting to carry out this same work with other data. For example, those of journals that practice open peer review. In other words, it would be interesting if we reached the same conclusions (or not) if we used more systematic and random data. This amounts to extracting on the one hand the reports of the publications and on the other hand, the citations received. Another limitation, which is difficult to overcome, is that this work is based on reports from accepted publications. It would be interesting to include in the analysis the length of reviewers' reports of rejected publications.

## References

Azar, O. H. (2007). The Slowdown in First-Response Times of Economics Journals: Can It Be Beneficial? *Economic Inquiry*, *45*(1), 179–187. https://doi.org/10.1111/j.1465-7295.2006.00032.x

Bilalli, B., Munir, R. F., & Abelló, A. (2020). A framework for assessing the peer review duration of journals: Case study in computer science. *Scientometrics*. https://doi.org/10.1007/s11192-020-03742-9

Björk, B.-C. (2018). Publishing speed and acceptance rates of open access megajournals. *Online Information Review*, *ahead-of-print*(ahead-of-print). https://doi.org/10.1108/OIR-04-2018-0151

Björk, B.-C., & Solomon, D. (2012). Open access versus subscription journals: A comparison of scientific impact. *BMC Medicine*, *10*(1), Article 1. https://doi.org/10.1186/1741-7015-10-73

Björk, B.-C., & Solomon, D. (2013). The publishing delay in scholarly peer-reviewed journals. *Journal of Informetrics*, *7*(4), 914–923. https://doi.org/10.1016/j.joi.2013.09.001

Boukacem-Zeghmouri, C., Rakotoary, S., & Bador, P. (2020). *La prédation dans le champ de la publication scientifique: Un objet de recherche révélateur des mutations de la communication scientifique ouverte*. https://hal.archives-ouvertes.fr/hal-02941731

Bravo, G., Grimaldo, F., López-Iñesta, E., Mehmani, B., & Squazzoni, F. (2019). The effect of publishing peer review reports on referee behavior in five scholarly journals. *Nature Communications*, *10*(1), Article 1. https://doi.org/10.1038/s41467-018-08250-2

Chataway, J., & Severin, A. (2020). Purposes of peer review: A qualitative study of stakeholder expectations and perceptions. *Learned Publishing*. https://discovery.ucl.ac.uk/id/eprint/10110911/

Cobey, K. D., Grudniewicz, A., Lalu, M. M., Rice, D. B., Raffoul, H., & Moher, D. (2019). Knowledge and motivations of researchers publishing in presumed predatory journals: A survey. *BMJ Open*, *9*(3), e026516. https://doi.org/10.1136/bmjopen-2018-026516

Cornelius, J. L. (2012). Reviewing the review process: Identifying sources of delay. *The Australasian Medical Journal*, *5*(1), 26–29. https://doi.org/10.4066/AMJ.2012.1165

Deming, W. E., & Stephan, F. F. (1940). On a Least Squares Adjustment of a Sampled Frequency Table When the Expected Marginal Totals are Known. *The Annals of Mathematical Statistics*, *11*(4), 427–444.

Deville, J.-C., Sarndal, C.-E., & Sautory, O. (1993). Generalized raking procedures in survey sampling. *Journal of the American Statistical Association*, *88*(423), 1013–1021.

Djuric, D. (2015). Penetrating the Omerta of Predatory Publishing: The Romanian Connection. *Science and Engineering Ethics*, *21*(1), 183–202. https://doi.org/10.1007/s11948-014-9521-4

Drvenica, I., Bravo, G., Vejmelka, L., Dekanski, A., & Nedić, O. (2019). Peer Review of Reviewers: The Author's Perspective. *Publications*, *7*(1), Article 1. https://doi.org/10.3390/publications7010001

Frandsen, T. F. (2019). Why do researchers decide to publish in questionable journals? A review of the literature. *Learned Publishing*, *32*(1), 57–62. https://doi.org/10.1002/leap.1214

Huisman, J., & Smits, J. (2017). Duration and quality of the peer review process: The author's perspective. *Scientometrics*, *113*(1), 633–650. https://doi.org/10.1007/s11192-017-2310-5

Kareiva, P., Marvier, M., West, S., & Hornisher, J. (2002). Slow-moving journals hinder conservation efforts. *Nature*, *420*(6911), Article 6911. https://doi.org/10.1038/420015a

Kolata, G. (2017, October 30). Many Academics Are Eager to Publish in Worthless Journals (Published 2017). *The New York Times*. https://www.nytimes.com/2017/10/30/science/predatory-journals-academics.html

Larivière, V. (2017). Croissance des revues savantes: De la connaissance et… du bruit. *Acfas*. https://www.acfas.ca/publications/magazine/2017/12/croissance-revues-savantes-entre-bruit-connaissances

Maddi, A., & Sapinho, D. (2022). Article processing charges, altmetrics and citation impact: Is there an economic rationale? *Scientometrics*. https://doi.org/10.1007/s11192-022-04284-y

Moizer, P. (2009). Publishing in accounting journals: A fair game? *Accounting, Organizations and Society*, *34*(2), 285–304. https://doi.org/10.1016/j.aos.2008.08.003

Pautasso, M., & Schäfer, H. (2010). Peer review delay and selectivity in ecology journals. *Scientometrics*, *2*(84), 307–315. https://doi.org/10.1007/s11192-009-0105-z

Pranić, S. M., Malički, M., Marušić, S. L., Mehmani, B., & Marušić, A. (2020). Is the quality of reviews reflected in editors' and authors' satisfaction with peer review? A cross-sectional study in 12 journals across four research fields. *Learned Publishing*, *n/a*(n/a), 1–11. https://doi.org/10.1002/leap.1344

Publons. (2018, February 26). *It's not the size that matters*. Publons. https://publons.com/blog/its-not-the-size-that-matters/

Radzvilas, M., De Pretis, F., Peden, W., Tortoli, D., & Osimani, B. (2022). Incentives for Research Effort: An Evolutionary Model of Publication Markets with Double-Blind and Open Review. *Computational Economics*. https://doi.org/10.1007/s10614-022-10250-w

Shen, C., & Björk, B.-C. (2015). 'Predatory' open access: A longitudinal study of article volumes and market characteristics. *BMC Medicine*, *13*(1), 230. https://doi.org/10.1186/s12916-015-0469-2

Shen, S., Rousseau, R., Wang, D., Zhu, D., Liu, H., & Liu, R. (2015). Editorial delay and its relation to subsequent citations: The journals Nature, Science and Cell. *Scientometrics*, *105*(3), 1867–1873. https://doi.org/10.1007/s11192-015-1592-8

Siler, K. (2020). Demarcating spectrums of predatory publishing: Economic and institutional sources of academic legitimacy. *Journal of the Association for Information Science and Technology*, *71*(11), 1386–1401. https://doi.org/10.1002/asi.24339

Toroser D, Carlson J, Robinson M, Gegner J, Girard V, Smette L, Nilsen J, & O'Kelly J. (2016). Factors impacting time to acceptance and publication for peer-reviewed publications. *Current Medical Research and Opinion*, *33*(7), 1183–1189. https://doi.org/10.1080/03007995.2016.1271778

Waltman, L., van Eck, N. J., van Leeuwen, T. N., Visser, M. S., & van Raan, A. F. J. (2011). Towards a new crown indicator: Some theoretical considerations. *Journal of Informetrics*, *5*(1), Article 1. https://doi.org/10.1016/j.joi.2010.08.001

# Appendices

## A.1. Structure of the Publons sample and of the WoS population, before and after adjustment

| Variable | Modalities | Number | % | Marginal sums on population | Marginal sums on population (%) | Publons' structure after adjustement (%) |
|---|---|---|---|---|---|---|
| Publication year | 2009 | 761 | 1.31 | 552 615 | 4.48 | 4.48 |
| | 2010 | 946 | 1.63 | 546 703 | 4.44 | 4.44 |
| | 2011 | 1 356 | 2.33 | 717 698 | 5.82 | 5.82 |
| | 2012 | 1 762 | 3.03 | 839 311 | 6.81 | 6.81 |
| | 2013 | 2 601 | 4.48 | 1 002 514 | 8.13 | 8.13 |
| | 2014 | 3 521 | 6.06 | 1 141 004 | 9.26 | 9.26 |
| | 2015 | 4 742 | 8.16 | 1 249 900 | 10.14 | 10.14 |
| | 2016 | 5 917 | 10.19 | 1 348 223 | 10.94 | 10.94 |
| | 2017 | 14 336 | 24.68 | 1 591 868 | 12.91 | 12.91 |
| | 2018 | 18 699 | 32.19 | 1 679 518 | 13.63 | 13.63 |
| | 2019 | 3 046 | 5.24 | 1 140 289 | 9.25 | 9.25 |
| | 2020 | 401 | 0.69 | 516 787 | 4.19 | 4.19 |
| | 2021 | 5 | 0.01 | 282 | 0.00 | 0.00 |
| Open acces | 1 | 25 177 | 43.34 | 3 532 995 | 28.66 | 28.66 |
| | 2 | 32 916 | 56.66 | 8 793 717 | 71.34 | 71.34 |
| Without financing | 1 | 15 284 | 26.31 | 5 681 665 | 46.09 | 46.09 |
| | 2 | 42 809 | 73.69 | 6 645 047 | 53.91 | 53.91 |
| One financing | 1 | 13 809 | 23.77 | 2 613 911 | 21.21 | 21.21 |
| | 2 | 44 284 | 76.23 | 9 712 801 | 78.79 | 78.79 |
| Two financing | 1 | 10 758 | 18.52 | 1 762 755 | 14.30 | 14.30 |
| | 2 | 47 335 | 81.48 | 10 563 957 | 85.70 | 85.70 |
| Three financing | 1 | 7 155 | 12.32 | 921 952 | 7.48 | 7.48 |
| | 2 | 50 938 | 87.68 | 11 404 760 | 92.52 | 92.52 |
| Four financing or more | 1 | 11 087 | 19.08 | 1 346 429 | 10.92 | 10.92 |
| | 2 | 47 006 | 80.92 | 10 980 283 | 89.08 | 89.08 |
| # Countries | 1 | 41 008 | 70.59 | 10 640 214 | 86.32 | 86.32 |
| | 2 | 12 778 | 22.00 | 1 485 416 | 12.05 | 12.05 |
| | 3 | 3 336 | 5.74 | 167 730 | 1.36 | 1.36 |
| | 4 | 971 | 1.67 | 33 352 | 0.27 | 0.27 |
| Total | | 58 093 | 100.0 | 12 326 712 | 100.0 | 100.0 |

## A.2. Structure of the Publons sample and of the WoS population, before and after adjustment*

| Variable - ERC Class | Modalities | Number | % | Marginal sums on population | Marginal sums on population (%) | Publons' structure after adjustement (%) |
|---|---|---|---|---|---|---|
| LS09 | 1 | 53 | 0.09 | 9 830 | 0.08 | 0.08 |
|  | 2 | 58 040 | 99.91 | 12 316 882 | 99.92 | 99.92 |
| LS1 | 1 | 6 255 | 10.77 | 695 214 | 5.64 | 5.64 |
|  | 2 | 51 838 | 89.23 | 11 631 498 | 94.36 | 94.36 |
| LS2 | 1 | 6 054 | 10.42 | 489 234 | 3.97 | 3.97 |
|  | 2 | 52 039 | 89.58 | 11 837 478 | 96.03 | 96.03 |
| LS3 | 1 | 1 855 | 3.19 | 129 674 | 1.05 | 1.05 |
|  | 2 | 56 238 | 96.81 | 12 197 038 | 98.95 | 98.95 |
| LS4 | 1 | 9 684 | 16.67 | 1 668 900 | 13.54 | 13.54 |
|  | 2 | 48 409 | 83.33 | 10 657 812 | 86.46 | 86.46 |
| LS5 | 1 | 4 350 | 7.49 | 555 326 | 4.51 | 4.51 |
|  | 2 | 53 743 | 92.51 | 11 771 386 | 95.49 | 95.49 |
| LS6 | 1 | 3 805 | 6.55 | 410 871 | 3.33 | 3.33 |
|  | 2 | 54 288 | 93.45 | 11 915 841 | 96.67 | 96.67 |
| LS7 | 1 | 17 372 | 29.90 | 2 824 978 | 22.92 | 22.92 |
|  | 2 | 40 721 | 70.10 | 9 501 734 | 77.08 | 77.08 |
| LS8 | 1 | 8 051 | 13.86 | 770 001 | 6.25 | 6.25 |
|  | 2 | 50 042 | 86.14 | 11 556 711 | 93.75 | 93.75 |
| LS9 | 1 | 8 065 | 13.88 | 1 202 239 | 9.75 | 9.75 |
|  | 2 | 50 028 | 86.12 | 11 124 473 | 90.25 | 90.25 |
| PE09 | 1 | 53 | 0.09 | 9 830 | 0.08 | 0.08 |
|  | 2 | 58 040 | 99.91 | 12 316 882 | 99.92 | 99.92 |
| PE1 | 1 | 2 029 | 3.49 | 326 962 | 2.65 | 2.65 |
|  | 2 | 56 064 | 96.51 | 11 999 750 | 97.35 | 97.35 |
| PE10 | 1 | 7 293 | 12.55 | 1 411 496 | 11.45 | 11.45 |
|  | 2 | 50 800 | 87.45 | 10 915 216 | 88.55 | 88.55 |
| PE11 | 1 | 5 288 | 9.10 | 1 227 477 | 9.96 | 9.96 |
|  | 2 | 52 805 | 90.90 | 11 099 235 | 90.04 | 90.04 |
| PE2 | 1 | 5 452 | 9.38 | 1 039 190 | 8.43 | 8.43 |
|  | 2 | 52 641 | 90.62 | 11 287 522 | 91.57 | 91.57 |
| PE3 | 1 | 3 419 | 5.89 | 659 856 | 5.35 | 5.35 |
|  | 2 | 54 674 | 94.11 | 11 666 856 | 94.65 | 94.65 |
| PE4 | 1 | 6 763 | 11.64 | 1 441 848 | 11.70 | 11.70 |
|  | 2 | 51 330 | 88.36 | 10 884 864 | 88.30 | 88.30 |
| PE5 | 1 | 5 917 | 10.19 | 1 294 056 | 10.50 | 10.50 |
|  | 2 | 52 176 | 89.81 | 11 032 656 | 89.50 | 89.50 |
| PE6 | 1 | 2 039 | 3.51 | 959 256 | 7.78 | 7.78 |
|  | 2 | 56 054 | 96.49 | 11 367 456 | 92.22 | 92.22 |
| PE7 | 1 | 3 302 | 5.68 | 1 477 664 | 11.99 | 11.99 |
|  | 2 | 54 791 | 94.32 | 10 849 048 | 88.01 | 88.01 |
| PE8 | 1 | 5 946 | 10.24 | 1 636 128 | 13.27 | 13.27 |
|  | 2 | 52 147 | 89.76 | 10 690 584 | 86.73 | 86.73 |
| PE9 | 1 | 3 221 | 5.54 | 652 903 | 5.30 | 5.30 |
|  | 2 | 54 872 | 94.46 | 11 673 809 | 94.70 | 94.70 |
| SH1 | 1 | 1 153 | 1.98 | 330 993 | 2.69 | 2.69 |
|  | 2 | 56 940 | 98.02 | 11 995 719 | 97.31 | 97.31 |
| SH2 | 1 | 516 | 0.89 | 95 174 | 0.77 | 0.77 |
|  | 2 | 57 577 | 99.11 | 12 231 538 | 99.23 | 99.23 |
| SH3 | 1 | 1 081 | 1.86 | 361 067 | 2.93 | 2.93 |
|  | 2 | 57 012 | 98.14 | 11 965 645 | 97.07 | 97.07 |
| SH4 | 1 | 2 313 | 3.98 | 363 528 | 2.95 | 2.95 |
|  | 2 | 55 780 | 96.02 | 11 963 184 | 97.05 | 97.05 |
| SH5 | 1 | 129 | 0.22 | 53 252 | 0.43 | 0.43 |
|  | 2 | 57 964 | 99.78 | 12 273 460 | 99.57 | 99.57 |
| SH6 | 1 | 97 | 0.17 | 9 486 | 0.08 | 0.08 |
|  | 2 | 57 996 | 99.83 | 12 317 226 | 99.92 | 99.92 |
| SH7 | 1 | 3 246 | 5.59 | 359 771 | 2.92 | 2.92 |
|  | 2 | 54 847 | 94.41 | 11 966 941 | 97.08 | 97.08 |
| Total |  | 58 093 | 100.0 | 12 326 712 | 100.0 | 100.0 |

*For the labels of the disciplines see: https://figshare.com/articles/dataset/OST_-_Classification_of_WoS_subject_categories_into_27_2_ERC_panels_/21707543